\documentstyle{mn}

\title{Structural properties of spherical galaxies
obeying \\ 
the S\'ersic $R^{1/n}$ law: a semi-analytical approach}
\author[Eduardo Simonneau and Francisco Prada]
       {Eduardo Simonneau$^1$ and Francisco Prada$^2$\thanks{
Email: prada@caha.es} \\
       $^1$Institut de Astrophysique, CNRS, 98bis, Bd. Arago,
           F-75014 Paris, France \\
       $^2$Centro Astron\'omico Hispano-Alem\'an, Apdo 511, 
E-04080, Almeria, Spain}
\date{Accepted 0000 December 00.
      Received 0000 December 00;
      in original form 0000 October 00}

\begin{document}

\maketitle


\begin{abstract}

By considering that the distribution of measured light along
any galactocentric radius of an elliptical galaxy can be
well represented by a $R^{1/n}$ S\'ersic law, we propose 
in this paper a ``discrete ordenate'' method, which,
for any value of $n$,  allows
an explicit expression  for the luminous
density, $\rho(r)$, to be found that can be numerically evaluated
to any required precision. Once we have this semi-analytical
expression for the spatial density, $\rho(r)$, the mass distribution,
 $M(r)$, the potential, $\phi(r)$, and the
velocity dispersions, $\sigma_s^2(r)$, in the space and 
in the observational plane, $\sigma_p^2(R)$, can
be  computed in a straightforward manner.

\end{abstract}


\section{Introduction}

    Analysis of the distribution of the light intensity
over the surface of an elliptical galaxy is fundamental
to determining its internal structure. From the beginning
it was pointed out that this intensity distribution
over a galactocentric radius---for simplicity along
the major axis of the observed ellipse---was similar for all
galaxies, as well as for the bulges of spiral galaxies.
 This distribution was firstly represented by the
Hubble--Reynolds law $I(R) \propto (r+r_o)^{-2}$
(Reynolds 1913; Hubble 1930),
and some time later by the ``universal'' de Vaucouleurs law. However, later observations 
(Davies et al. 1988; Caon, Capaccioli \& D'Onofrio 1993; Andreakis, Peletier 
\& Balcells 1995;
Graham et al. (1996); Binggeli \& Jergen 1998) did 
show some discrepancies with the $R^{1/4}$ law, and led to
a more general form with $R^{1/n}$, generally with $n>1$ (S\'ersic 1968).
This bi-parametric expression accounts much better for the observed
intensity profiles of elliptical galaxies and bulges
of spiral galaxies.

 In this paper we present a careful study to determine the 3D
spatial distribution of
emitting sources from any observed $R^{1/n}$ S\'ersic profile, and
via a mass-to-luminosity relation we derive the corresponding dynamical
properties.

\section[]{The Sersic profile}

   For an elliptical galaxy the ``light profile'' which describes the 
light intensity distribution over any galactocentric radius
 $R(\alpha)$ corresponding to the galactic
 longitude $\alpha$  can 
be well approached by the S\'ersic profile, 

\begin{equation}
I(R(\alpha))=I(0) \exp [-(R(\alpha)/R_{\rm {o}}(\alpha))^{{1 \over n}}],
\end{equation}
 where the value of $n$ is the same for all the galactocentric
radius $R(\alpha)$. The scale length, $R_o(\alpha)$, will 
have different values for different $\alpha$ according with 
the elliptical shape of the galaxy. If $a$ and $b$ are, respectively,
the major and minor semi-axis of a given isophote ($\epsilon=1-b/a$ is 
the ellipticity) and $R_o \equiv R_o(0)$ is the scale length of the 
light profile along the major semi-axis, we then have,

\begin{equation}
 R_o(\alpha)^2=R_o^2 { (1-\epsilon)^2 \over \sin^2\alpha + (1-\epsilon)^2 \cos^2\alpha }.
\end{equation}

Equations (1) and (2), with $n$ independent of $\alpha$, indicate
that all the isophotes (curves with the same value of the
light intensity, $I$) are ellipses with the same ellipticity, $\epsilon$.

 In general, the observational intensity profiles are measured 
along the major axis of the galaxy, where $\alpha=0$ 
and therefore we note $I(R) \equiv I(R(0))$.  However,
 other parameters have been proposed for $R$ (radius) for identifying 
each observed isophote and measuring
the intensity profile. For example, if $a_k$ and $b_k$ are
the major and minor
semi-axes of each isophote, respectively, instead of $R=a_k$ we can use 
the value of $b_k$, the mean
value  $(a_k+b_k)/2$ or the geometric mean ($\sqrt{a_kb_k}$). 
However,
each of these options leads to the same conclusions. We might
have only small differences from the practical point of view. Therefore,
purely for sake of simplicity we carry on with our work using the semi-major 
axis as the ``radius parameter'', $R$, to measure the distribution of 
luminous intensity over the galaxy.

 The total energy or luminosity enclosed by an isophote defined
by the value $R$ of its major axis is given by,

\begin{equation}
   L (R)={1 \over 2} \int_{0}^{2\pi} {\rm{d}}\alpha
 \int_{0}^{R(\alpha)} R^\prime(\alpha) I(R^\prime(\alpha)) 
{\rm{d}}R^\prime(\alpha),
\end{equation}
 $R(\alpha)$ being the value of the radius vector
of the points of the ellipse-isophote defined
by $R$ that corresponds
to the galactocentric longitude $\alpha$. If the brightness distribution 
follows the S\'ersic law, Equation (1),
and taking into account that for the isophote defined by $R$ we 
have $R(\alpha)/R_o(\alpha)=R/R_o$, then Equation (3) reduces to

\begin{equation}
L(R)=I(0) R_o^2 \pi (1- \epsilon) 2n  
\gamma(2n,(R/R_o)^{{1 \over n}}),
\end{equation}
 $\gamma(2n,(R/R_o)^{1/n})$ being the incomplete gamma function corresponding 
to the gamma function $\Gamma(2n)$ for the value $(R/R_o)^{1/n}$.

The total observed luminosity is given by Equation (4) when 
$R\rightarrow \infty$:

\begin{equation}
   L=I(0) R_o^2 \pi (1-\epsilon) 2n  \Gamma(2n).
\end{equation}

But instead of using the natural scale length $R_o$
 to describe a profile $I(R)$  measured 
along the semi-major axis of the
isophotes,  the value $R_e$ of
$R$, for which the corresponding isophote encloses the half of
the total luminosity is customarily used. Obviously, the other half corresponds
to the luminosity over the isophotes with $R>R_e$. This definition
of $R_e$ can be written as 

\begin{equation}
  {1 \over 2} \Gamma (2n) = \gamma (2n,(R_e/R_o)^{1 \over n}).
\end{equation}

This equation gives a value of $R_e/R_o$ for each $n$. If we define
$k=(R_e/R_o)^{1/n}$, the values of $k$ that satisfy Equation (6)
can be easily computed for any value of $n$. For $n \geq 1$, they
can be obtained (with an error smaller than 0.1$\%$) by the 
relation $k=2n-0.324$ (see Ciotti 1991).
   
The use of the variable $k$ and the characteristic radius $R_e$ allow us to
write the S\'ersic profile, Equation (1), in its usual form (see Fig. 1):

\begin{equation}
I(R)=I(0) \exp [-k(R/R_{\rm {e}})^{{1 \over n}}].
\end{equation}

Likewise, the luminosity distribution in Equation (4) becomes,

\begin{equation}
L(R)=I(0) R_e^2 \pi (1-\epsilon) {2n \over k^{2n}} \gamma(2n,(R/R_e)^{1/n}),
\end{equation}
and the total luminosity is,

\begin{equation}
L=I(0) R_e^2 \pi (1-\epsilon) {2n \over k^{2n}} \Gamma(2n).
\end{equation}

In Fig. 1 we plot, together with the observed intensity profile $I(R)$,
the normalized observed luminosity, $L(R)/L$, for different values of $n$.

\section{The Abel inversion for the Sersic profiles} 

  The problem that we want to solve is to determine
the galactic density of luminous sources which can explain the
observed intensity profile, $I(R)$, given by
Equations (1) or (7). This intensity at a given
position is originated for by the emission of all the stars that
are located along the line of sight. For the
case in which the emitting sources of the galaxy have
a spherical distribution of density, $\rho_{L}(r)$,
the observed isophotes will be circular and
the corresponding intensity will be given by

\begin{equation}
   I(R)= \int_{R}^{\infty} \rho_{L}(r) {2r{\rm{d}}r \over \sqrt{r^2-R^2}} ,
\end{equation}  
where $R$ is the radius of each observed isophote. The problem that we 
have to solve is to determine the density of luminosity, $\rho_{L}(r)$, for 
a known intensity profile, $I(R)$. In mathematical terms the
problem consists in the inversion of the Abell integral transform 
with an infinite interval (Equation [10]).

 If the distribution of emitting sources is not spherical, as
happens in the case of elliptical galaxies, the problem is not
so simple. Here we can admit that the density of sources
$\rho_{L}$ is constant over concentric and coaxial ellipsoidal
surfaces which can be bi- or triaxial. Let $d$ be the parameter
whose value characterizes each  of the ellipsoidal shells in the 
space. It 
can be the value of one of its three semi-axis, a mean 
(linear or quadratic)
value of these axes, or any other parameter with linear dimensions. Then
the density of luminous sources, $\rho_{L}(d)$, will be a function 
of $d$. If these ellipsoidal shells are homologous (i.e. if the
ratio between their semi-axes are the same for all of them) the
corresponding isophotes over any observation plane are concentric
and coaxial homologous ellipses (all have the same ellipticity). 
Taking $D$ as a parameter to describe the observed
isophotes ($D$ can be the semi-major axis or any
other linear dimension of these isophotes 
as we discuss in Section 2), we can write an equation
similar to Equation (10) (see Lindblad 1956), which 
relates the intensity, $I(D)$, of each isophote with the density of emitting sources,
$\rho_{L}(d)$. Now $d$ plays the same role as $r$ in the case 
of a spherical distribution of sources, but $d$ is
 no longer
a vector position in the space. The choice of the
parameter $D$ to characterize each of the observed
isophotes and to describe the distribution of
the intensity, $I(D)$, over these depends on
the choice of the parameter $d$ to describe
the structure and emission density, $\rho_{L}(d)$,
of each ellipsoidal shell that forms the galaxy.
To find the relation between $D$ and $d$ we need
a detailed geometrical description of the shape,
position and projection of these ellipsoidal surfaces
(see Simonneau, Varela \& Mu\~noz-Tu\~n\'on 1998).

  However the main goal of this work is to describe in a semi-analytical
way, some mathematical properties of a density 
$\rho$$_{L}$($r$) obtained
from an inversion of the Abel integral transform (Equation [10])
of an observed radial profile, $I$($R$), which follows the S\'ersic law
(Equation [7]). These mathematical properties are the same 
for the case
of a spherical galaxy where the emitting shells
are defined by $r$ and the circular isophotes
are defined by its radius, $R$, which, for the
case of an ellipsoidal galaxy where the
emitting shells are defined by the 
parameter $d$ and the elliptical
isophotes, are defined by another parameter, $D$. 
In this paper, then, we shall  consider only the case of
the spherical galaxies by means of Equation (10). But 
we would
like to emphasize that many of the conclusions presented
in this paper can be applied, with minor changes, to the
case of elliptical galaxies, once we find out how to define 
 the $d$ and $D$ parameters correctly.

 In terms of the new variables $z \equiv (R/R_e)$ and $s\equiv (r/R_e)$,
the theoretical inversion of the Abel integral transform 
(Tricomi 1985), equation (10), leads to

\begin{equation}
  R_e \rho_{L}(s)= - {1 \over \pi} \int_{s}^{\infty}  
  { {\rm{d}} I(z) \over {\rm{d}}z} { {\rm{d}}z \over \sqrt{z^2-s^2} },
\end{equation}
where $I(z)$ is given by the Equation (7). It will then be

\begin{equation}
  \rho_{L}(s)= {1 \over \pi} {k \over n} {I(0) \over R_e} \int_{s}^{\infty}  
{  \exp[-kz^{ {1 \over n} }] \, z^{ {1 \over n}-1} \over \sqrt{z^2-s^2} } 
{\rm{d}}z,
\end{equation}
or otherwise

\begin{equation}
  \rho_{L}(s)= {k \over \pi} {I(0) \over R_e} 
s^{ {1 \over n}-1 } 
 \int_{1}^{\infty}  {  \exp[-ks^{{1 \over n}}t] 
 \over \sqrt{t^{2n}-1} } {\rm{d}}t.
\end{equation}

The argument in the integral of  Equation (13) is singular
at $t=1$, although it is integrable for any value of $n$. Only
when $s=0$ does the integral diverge for $n<1$ because of
the infinity in the upper limit. But for these
values, because of the multiplicative factor $s^{{1/n}-1}$,
$\rho_{L}(s)$ becomes zero for $s=0$. When
$s$ increases, $\rho_{L}(s)$ also increases  until some critical
value and then decreases monotonically. This means that
some layers are more dense than the internal ones, and that consequently
the system is gravitationally unstable. Hence, models with
$n<1$ are not adequate for representing stable galaxies.

 When $n=1$ the intensity profile, $I(R)$, takes an exponential
law and the integral in Equation (13) is the $K_o(x)$ modified
Bessel function, which has a logarithmic discontinuity
for $x=0$ (see Abramowitz \& Stegun 1968, pages 374--376). Hence,
for $n=1$ the spatial density becomes:

\begin{equation}
  \rho_{L}(s)= {k \over \pi} {I(0) \over R_e} K_o(k s).
\end{equation}

 For $n>1$ the integration of Equation (13) is not  possible analytically,
except for $s=0$, where

\begin{equation}
  \int_{1}^{\infty} { {\rm{d}}t  \over \sqrt{t^{2n}-1}} = 
 {1 \over 2n} B({1 \over 2}, {n-1 \over 2n}) 
\end{equation}
and $B(z,w)$ is the complete beta (Euler)
function (see Abramowitz \& Stegun 1968, page 258).

However, the form of the integral in Equation (13) has been used
by many authors to obtain selected approximations
for $\rho_{L}(s)$. First,
for the de Vaucouleurs case $n=4$, by Poveda, Iturriaga \& Orozco (1960),
Mellier \& Mathez (1987), and for the S\'ersic profiles
(Gerbal et al. 1997). Also, numerical computations of
$\rho_{L}(s)$ have been carried out by Young (1976)
for the de Vaucouleurs case
and by Ciotti (1991) and Graham \& Colles (1997) for the S\'ersic 
cases, respectively. In this article we propose 
an analytical expression for $\rho_{L}(s)$ that allows an easy 
computation of the mass and gravitational potential
to any required precision.

  As the argument of the integral
in Equation (13) can be integrated for any
value of $s$, even considering the singularity for $t=1$, it seems
possible to write this integral with a new variable, $x$,
such as its argument does not show any discontinuity. We
propose $t=1/(1-x^2)^{{1/(n-1)}}$. Equation (13) for
the density becomes

\begin{equation}
  \rho_{L}(s)= {k \over \pi} {I(0) \over R_e} 
{2 \over n-1} {1 \over s^{ { n-1 \over n } } }
 \int_{0}^{1}  {  \exp[-ks^{{1 \over n}}(1-x^2)^{-{1 \over n-1}}] 
 \over \sqrt{1-(1-x^2)^{{2n \over n-1}}} } \,\, x {\rm{d}}x.
\end{equation}

We can then perform  this integral formally by means of a 
Gaussian numerical integration to get

\begin{equation}
  \rho_{L}(s)= {k \over \pi} {I(0) \over R_e} {2 \over n-1}
{1 \over s^{ { n-1 \over n } } } 
 \sum_{j=1}^{N_{ap}}
\rho_j \exp[-\lambda_j k s^{{1 \over n}}],
\end{equation}
where

\begin{equation}
  \lambda_j= { 1 \over (1-x_j^2)^{ {1 \over n-1} } }, 
\end{equation}

\begin{equation}
  \rho_j= w_j { x_j \over \sqrt{1-(1-x_j^2)^{ {2n \over n-1} } } }
\end{equation}
and $x_j$ and $w_j$ ($j=1,2,...,N_{ap}$) are the abcissae and the 
integration weights, respectively, in the interval $(0,1)$. $N_{ap}$ is
the order of approximation, i.e. number of abscissae and weights needed
to compute  the integral in Equation (13) numerically. The standard 
tables of abcissae, 
$\bar{x}_j$, and integration weights, $\bar{w}_j$, correspond to
the interval $(-1,+1)$ (Abramowitz \& Stegun 1968, pages 916--919). Therefore,
the relation of the values of $x_j$ and $w_j$  for the interval
(0,1) are $x_j=(1+\bar{x}_j)/2$ and $w_j$=$\bar{w}_j$/2. In summary,
by selecting different sets of abcissae, $x_j$, and
integration weights, $w_j$, we can compute $\lambda_j$ and
the coefficients $\rho_j$ by Equations (18) and (19), respectively,
and   determine
to any required precision the luminous
density $\rho_{L}(s)$ from Equation (17). In Table 1, we have listed the abscissae
and weights for the $N_{ap}=1,2$ and 5 cases that we  discuss 
below.

\begin{table*}
\begin{minipage}{80mm}
\caption{Abscissae, $x_j$, and weights, $w_j$,  for the
Gaussian  
integration in the interval (0,1) for
three different approximations,  
$N_{ap}$=1,2 and 5.}
\begin{tabular}{@{}llrrlr@{}}
	  & $x_j$    & $w_j$    \\
          &          &          \\
$N_{ap}=1$    & 0.5      & 1.       \\
          &          &          \\
$N_{ap}=2$    & 0.211325 & 0.5      \\
          & 0.788675 & 0.5      \\
          &          &          \\
$N_{ap}=5$    & 0.046910 & 0.118464 \\
          & 0.230765 & 0.239314 \\
          & 0.5      & 0.284444 \\
          & 0.769235 & 0.239314 \\
          & 0.953090 & 0.118464 \\  
\end{tabular}
\end{minipage}
\end{table*}

In the first place, we must point out that there are few
characteristic quantities in relation to the 
distribution  $\rho_{L}(s)$
that can be determined theoretically, i.e. independently
of the numerical integrations. One of these is the
integral in Equation (15), which determines the asymptotical
behaviour of $\rho_{L}(s)$ when $s\rightarrow 0$. From Equations (15)
to (17) it is necessary that

\begin{equation}
 {4n \over n-1} \sum_{j=1}^{N_{ap}} \rho_j \sim  B({1\over2},{n-1\over2n}),
\end{equation}
where the coefficients $\rho_j$ are those given by 
Equation (19). 
The other characteristic quantities
are the different moments of the density distribution, $\rho_{L}(r)$,
i.e.

\begin{equation}
\int_{0}^{\infty}  \rho_{L}(r) r^{m} {\rm{d}}r.
\end{equation}

The first one ($m=0$) leads to the synthetic value of $I(0)$. To
satisfy Equation (10) for $R=0$ with a density law as in
Equation (17) it is necessary that

\begin{equation}
 \sum_{j=1}^{N_{ap}} { \rho_j \over \lambda_j } \sim
 {\pi \over 2} {n-1 \over 2n}.
\end{equation}

The second one ($m=1$) leads to the value of the gravitational
potential at the centre of the system ($r=0$). This value,
$\phi(0)$, is known a priori (see bellow Equation [31]). To reach this
theoretical value with a density law, as in Equation (17),
is necessary that

\begin{equation}
 \sum_{j=1}^{N_{ap}} { \rho_j \over {\lambda_j^{n+1}} } \sim
 {n-1 \over 2n}.
\end{equation}

The third one ($m=2$) leads to the value of total luminosity,
$L$. To have the same theoretical value as in Equation (9) it is
necessary that

\begin{equation}
 \sum_{j=1}^{N_{ap}} { \rho_j \over {\lambda_j^{2n+1}} } \sim
 {\pi \over 4} {n-1 \over 2n}.
\end{equation}

Thus, to test as a first step the accuracy of our approximation
for $\rho_{L}(s)$ in Equation (17), we have computed
 the left-hand side of  Equations (20),(22), (23) and (24)
numerically; and in
Tables 2, 3, 4 and 5
we have compared these with 
the theoretical values of their right-hand side together for the
different approximations $N_{ap}$=1, 2, and 5,
and for different values of $n$ (2,3,4,...,10). As  can be seen,
the approximation $N_{ap}=5$ leads to results of  high quality. 
According to 
the results we have that for $N_{ap}=10$,20 and 40, the 
variation is insignificant with respect to the results
for $N_{ap}=5$.
 
We now discuss the accuracy of each approximation 
($N_{ap}$=1,2,5,10,20,40) over  the entire radial density 
distribution, $\rho(s)$. As a reference we take the
density $\rho_{40}(s)$ calculated with $N_{ap}=40$. The relative
variation between this and $\rho_{20}(s)$, calculated with
the approximation $N_{ap}=20$ is always smaller than $0.01\%$, i.e.
negligible to  the first four significant figures.
 This
difference increases by one order of magnitude for the 
$N_{ap}=10$ approximation. The relative difference between
$\rho_{10}(s)$ and $\rho_{40}(s)$ is always smaller than
$0.1\%$. For the case of $N_{ap}=5$ the difference 
between $\rho_{5}(s)$ and $\rho_{40}(s)$ never reaches  $1\%$.
For the case of $N_{ap}=2$ we can get differences of around 
$10\%$
only for low values of $n$ (less that 5), and for large
values of $s$ (greater than 10). This relative difference
increases significantly for $N_{ap}=1$. So, if we consider as not
significant the difference between the  $N_{ap}=40$ and
$N_{ap}=20$ cases, we can take as the errors of each approximation
the differences given above. In general,  
$N_{ap}=2$ is a good approximation for getting a complete 
 first-order description. However, concerning the
quality of the first approximations ($N_{ap}$=1,2)
 we cannot consider  the Gaussian numerical
integration as the best method for obtaining the values of
$\rho_j$,$\lambda_j$. These values must 
be obtained directly from
Equations (20), (22), (23) and (24) in such a way that,
even if between these first orders the computed $\rho_{L}(s)$
and the correct solution there are any differences, we
are sure that the most important parameters of the problem
are computed exactly.

 For most  applications where  great accuracy is not
  needed, the approximation  $N_{ap}=5$ can be sufficient. In
   case of much higher accuracy the $N_{ap}=10$ and
$N_{ap}=20$ approximations can satisfy any requirement.

In Fig. 2, we plot $\rho_{L}(s)$ as a function of $s=r/R_e$
for different values of $n$.

\begin{table*}

\begin{minipage}{80mm}
\caption{Asymptotic coefficient of $\rho_{L}(s)$ for 
$s\rightarrow 0$,  Equation (20). We show the exact value of 
$B({1\over2},{n-1\over2n})$   and the corresponding 
computed values  ${4n \over n-1} \sum_{j=1}^{N_{ap}} \rho_j$, 
Equation (20), with $N_{ap}$=1,2,5.}
\begin{tabular}{@{}llrrlr@{}}
     &         \multicolumn{3}{c}
{ }\\
   n & $B({1\over2},{n-1\over2n})$ & $N_{ap}=1$ & $N_{ap}=2$ &
$N_{ap}=5$   \\
   &        &        &        &                         \\
2  & 5.2441151 & 4.837945 & 5.255770 & 5.244083         \\
3  & 4.2065460 & 3.945576 & 4.204081 & 4.206550         \\
4  & 3.8558066 & 3.643523 & 3.850307 & 3.855832         \\
5  & 3.6790940 & 3.490923 & 3.672538 & 3.679124         \\
6  & 3.5725536 & 3.398728 & 3.565529 & 3.572583         \\
7  & 3.5012896 & 3.336962 & 3.494024 & 3.501318         \\
8  & 3.4502622 & 3.292684 & 3.442860 & 3.450289         \\
9  & 3.4119198 & 3.259382 & 3.404436 & 3.411943         \\
10 & 3.3820539 & 3.233423 & 3.374518 & 3.382076         \\  
\end{tabular}
\end{minipage}
\end{table*}

\begin{table*}

\begin{minipage}{80mm}
\caption{Synthesis of the observed central  intensity $I(0)$. 
We show the theoretical value of ${\pi \over 2}{n-1\over2n}$ 
together with the   corresponding numerical computation 
of $\sum_{j=1}^{N_{ap}} \rho_j/\lambda_j$,  Equation (22),
with $N_{ap}$=1,2,5.}
\begin{tabular}{@{}llrrlr@{}}
     &         \multicolumn{3}{c}
{ }\\
   n & ${\pi \over 2}{n-1\over2n}$ & $N_{ap}=1$ &  $N_{ap}=2$ &
  $N_{ap}=5$  \\
   &        &        &        &                         \\
2  & 0.3926991 & 0.453557 & 0.397601 &  0.392702        \\
3  & 0.5235988 & 0.569495 & 0.537842 &  0.524518        \\
4  & 0.5890502 & 0.620693 & 0.603678 &  0.590514        \\
5  & 0.6283185 & 0.649734 & 0.641712 &  0.629974        \\
6  & 0.6544985 & 0.668478 & 0.666468 &  0.656182        \\
7  & 0.6731984 & 0.681587 & 0.683865 &  0.674842        \\
8  & 0.6872234 & 0.691273 & 0.696759 &  0.688799        \\
9  & 0.6981317 & 0.698723 & 0.706698 &  0.699631        \\
10 & 0.7068726 & 0.704633 & 0.714593 &  0.708280        \\  
\end{tabular}
\end{minipage}
\end{table*}

\begin{table*}

\begin{minipage}{80mm}
\caption{Synthesis of the theoretical  gravitational potential
at $r=0$. We show the theoretical value of ${n-1\over2n}$ 
together with  the corresponding numerical computation 
of $\sum_{j=1}^{N_{ap}} \rho_j/\lambda_j^{n+1}$,  Equation (23),
with $N_{ap}$=1,2,5.}
\begin{tabular}{@{}llrrlr@{}}
     &     \multicolumn{3}{c}
{ }\\
   n & ${n-1\over2n}$ & $N_{ap}=1$ &  $N_{ap}=2$ &
  $N_{ap}=5$  \\
   &        &        &        &                         \\
2  & 0.25      & 0.255126 & 0.246949 & 0.249997         \\
3  & 0.3333333 & 0.369898 & 0.327389 & 0.333356         \\
4  & 0.375     & 0.422952 & 0.370122 & 0.374965         \\
5  & 0.4       & 0.453484 & 0.396269 & 0.399942         \\
6  & 0.4166667 & 0.473327 & 0.413849 & 0.416597         \\
7  & 0.4285714 & 0.487259 & 0.426461 & 0.428498         \\
8  & 0.4375    & 0.497580 & 0.435943 & 0.437496         \\
9  & 0.4444444 & 0.505533 & 0.443330 & 0.444371         \\
10 & 0.45      & 0.511849 & 0.449245 & 0.449928         \\  
\end{tabular}
\end{minipage}
\end{table*}

\begin{table*}

\begin{minipage}{80mm}
\caption{Synthesis of the theoretical value of the total mass.
We show the theoretical value of ${\pi \over 4}{n-1\over2n}$ 
together with the  corresponding numerical computation 
of $\sum_{j=1}^{N_{ap}} \rho_j/\lambda_j^{2n+1}$,  Equation (24),
with $N_{ap}=1,2,5$.}
\begin{tabular}{@{}llrrlr@{}}
     &         \multicolumn{3}{c}
{ }\\
   n & ${\pi\over4}{n-1\over2n}$ & $N_{ap}=1$ &  $N_{ap}=2$ &
  $N_{ap}=5$ \\
   &        &        &        &                         \\
2  & 0.1963495 & 0.143508 & 0.208820 & 0.196357         \\
3  & 0.2617994 & 0.240256 & 0.265076 & 0.261793         \\
4  & 0.2945431 & 0.288208 & 0.294168 & 0.294525         \\
5  & 0.3141592 & 0.316511 & 0.312086 & 0.314166         \\
6  & 0.3272492 & 0.335146 & 0.324230 & 0.327259         \\
7  & 0.3365992 & 0.348335 & 0.333001 & 0.336611         \\
8  & 0.3436117 & 0.358159 & 0.339631 & 0.343623         \\
9  & 0.3490659 & 0.365757 & 0.344818 & 0.349077         \\
10 & 0.3534292 & 0.371810 & 0.348987 & 0.353441         \\  
\end{tabular}
\end{minipage}
\end{table*}

\section{Dynamical properties for the Sersic profile}

Once we have the density of luminous sources,  $\rho_{L}(s)$, we can study
 some other quantities in relation to the dynamical state of
the galaxy.

\subsection{The mass distribution}

 The density $\rho_{L}(s)$ in Equation (17) refers to the density of
luminous sources. In cases in which the mass-to-luminosity ratio,
$\Upsilon$, is the same in all the points of the galaxy, the total
mass, $M$, will be given by  Equation (9) multiplied by
this ratio $\Upsilon$. Then, we can use $\rho(s)$ for the mass
density distribution with the form

\begin{equation}
\rho(s) = \rho_o \bar{\rho}(s),
\end{equation}
where (with $M=\Upsilon L$),

\begin{equation}
  \rho_o= \Upsilon {L \over \pi^2 R_e^3} {2 \over n-1}
{k^{2n+1} \over \Gamma(2n+1) } 
\end{equation}
and 

\begin{equation}
  \bar{\rho}(s)= {1 \over s^{ { n-1 \over n } } } 
 \sum_{j=1}^{N_{ap}} \rho_j \exp[-\lambda_j k s^{{1 \over n}}].
\end{equation}

The corresponding mass distribution, $M(s)$,

\begin{equation}
  M(s)= R_e^3 \int_{0}^{s} 4 \pi s'^2 \rho(s') {\rm{d}}s,
\end{equation}
is given by

\begin{equation}
  {M(s) \over M} = {4 \over \pi (n-1) \Gamma(2n)} 
\sum_{j=1}^{N_{ap}} {\rho_j \over \lambda_j^{2n+1} }
 \gamma(2n+1,\lambda_j k s^{{1 \over n}}),
\end{equation}
where $\gamma(2n+1,\lambda_j k s^{{1 \over n}})$ is the incomplete
gamma function corresponding to the gamma function $\Gamma(2n+1)$
for the value $\lambda_j k s^{{1 \over n}}$. Because the Gaussian
values of $\lambda_j$ and $\rho_j$ satisfy 
Equation (24)  very accurately 
we are sure about the correct normalization in the 
$M(s)$ distribution.

The normalized mass distribution, $M(s)/M$, is shown in Fig. 2 for the
different values of $n$. For $n=1$ the distribution of mass
has been computed directly from the corresponding density given
by Equation (14).

\subsection{The gravitational potential}

 For a spherical system with density distribution $\rho(s)$
(Equation [25]), the 
gravitational potential is given by

\begin{equation}
\phi(s)= - 4 \pi G R_e^2 \rho_o ( {1 \over s} 
\int_{0}^{s} \bar{\rho}(s') s'^2 {\rm{d}}s'  +
\int_{s}^{\infty} \bar{\rho}(s') s'^2 {\rm{d}}s' )l,
\end{equation}
where, for the S\'ersic law, $\rho_o$ and $\bar{\rho}(s)$ are given by
Equations (26) and (27), respectively.

 At the centre, for $s=0$, we have

\begin{equation}
\phi(0)= - ({\Upsilon G L \over R_e }) {2 \over \pi}
k^n  { \Gamma(n) \over \Gamma(2n) },
\end{equation}
which is the correct value known a priori (see Ciotti 1991).

 For the potential we then have

\begin{equation}
\phi(s)= {2n \phi(0) \over (n-1) \Gamma(n+1) }
\sum_{j=1}^{N_ap} {\rho_j \over \lambda_j^{n+1}} 
\gamma(n+1,\lambda_j k s^{ {1 \over n} }) -
({\Upsilon G L \over R_e }) {1 \over s} { M(s) \over M }, 
\end{equation}
where $M(s)/M$ is given by  Equation (29). As before, the accuracy
of the sum $\sum_{j=1}^{N_ap} \rho_j/\lambda_j^{n+1}$ in Equation (23)
leads to the correct value for $\phi(0)$ in Equation (32).

Curves for the potential $\phi(s)$ for different values of $n$
($n=1,2,...,10$) are shown in Figure 3.

\subsection{The total potential energy}

With the above expression for $\rho(s)$, $M(s)$ and $\phi(s)$
it is very simple to compute the total potential energy,

\begin{equation}
W= -{1\over2} \int_{0}^{\infty} 4 \pi r^2 \rho(r) \phi(r) {\rm{d}}r,
\end{equation}
to get

\begin{equation}
W= \Upsilon^2 {G L^2 \over R_e} w^2.
\end{equation}

 The values of the parameter $w^2$, computed
numerically from Equation (33), are given in Table 6 for the different
values of $n$.

\begin{table*}

\begin{minipage}{80mm}
\caption{The potential energy, $w^2$, measured by taking 
($\Upsilon^2 GL^2/R_e$) as unity in Equation (34) 
for different values of $n$.}
\begin{tabular}{@{}llrrlr@{}}
$n$ & $w^2$   \\
   &     \\
1  & 0.31426    \\
2  & 0.30973    \\
3  & 0.31856    \\
4  & 0.33615    \\
5  & 0.35983    \\
6  & 0.38897    \\
7  & 0.42349    \\
8  & 0.46363    \\
9  & 0.50981    \\
10 & 0.56260    \\  
\end{tabular}
\end{minipage}
\end{table*}

The ratio $R_e/w^2$ can be considered as a mean gravitational
radius, $r_{\rm G}$, such that $W=\Upsilon^2GL^2/r_G$. That is, in some
way $r_{\rm G} \equiv R_e/w^2$ accounts for the mass concentration degree 
in the inner parts of the system. But from another point of view we can 
interpret ($\Upsilon$GL$/$$R_e$)$w^2$
as a quadratic mean velocity such that the corresponding
kinetic energy, $T=(1/2)M$($\Upsilon$GL$/$$R_e$)$w^2$, satisfies the virial 
theorem, $W=2T$. We should now examine the possibility of
measuring this quadratic mean velocity.

In stationary spherical galaxies we can assume that there are
no organized motions of the stellar populations as a whole, i.e. rotation,
expansion or contraction; hence at each point in the
galaxy we will have a quadratic mean velocity  
 corresponding to the local velocity dispersion 
$\langle$ $\bmath{\sigma_s}$$^2(r)$ $\rangle$.
Indeed this one is a double mean. First, it  is a local
quadratic mean of the three components of the velocity dispersion
$\bmath{\sigma_s}(r)$$ \equiv (\sigma_r(r),\sigma_{\phi}(r),\sigma_{\theta}(r))$.
However, to be able to apply the virial theorem we need the
necessary condition of quasi-isotropy in the velocity dispersion
tensor. In
practice we can admit  isotropy, i.e. 
$\sigma_r(r)=\sigma_{\phi}(r)=\sigma_{\theta}(r)\equiv\sigma_s(r)$,
to be able to apply the virial theorem with a quadratic mean velocity:

\begin{equation}
\langle \bmath{\sigma_s}^2(r) \rangle = 3 { \int_{0}^{\infty} \sigma_s(r) 
r^2 \rho(r) {\rm{d}}r  \over  \int_{0}^{\infty} r^2 \rho(r)
{\rm{d}}r } = 3 \sigma_s^2
\end{equation}

If observations of this quadratic mean 
velocity (dispersion velocity) are available,
and if we admit a ``S\'ersic model'' for the galaxy  and consequently  
 compute, for 
any observed $I(R)$ profile, the potential energy (Equations [34]; 
see Table 6 for the factor $w^2$ corresponding to each $n$), we could 
write the virial theorem in the form

\begin{equation}
M 3 \sigma_s^2  = \Upsilon^2 ({GL \over R_e}) L w^2,
\end{equation}
where the total luminosity, $L$, is another measured quantity. Therefore,
we can estimate the value of the mass-to-luminosity ratio, $\Upsilon$. This
method for the determination of the masses of  elliptical galaxies
was proposed, for the case $n=4$ (the de Vaucouleurs law) by Poveda (1958).

However, we do not have access to measurements of the 
velocity dispersion, $\sigma_s(r)$, in the space. We can 
measure only the velocity dispersion, $\sigma_p(R)$, in the
observation plane. Theoretically, these measurements can be
represented by

\begin{equation}
I(R) \sigma_p^2(R) = \int_{R}^{\infty} \sigma_s^2(r) \rho_{L}(r)
{2r \over \sqrt{r^2-R^2} } {\rm{d}}r,
\end{equation} 
which corresponds to the integral along the line of sight
of the spatial velocity dispersion, $\sigma_s^2(r)$, weighted
with the density. The quadratic mean value of 
$\sigma_p^2(R)$ is

\begin{equation}
\langle \sigma_p^2 \rangle = { \int_{0}^{\infty} \sigma_p^2(R) 
 {\rm{d}}L(R)  \over  L } = \sigma_p^2.
\end{equation}

 From Equations (35) to (37) and from the definition of $I(R)$
in Equation (10), the condition that $\sigma_p^2 = \sigma_s^2$
is satisfied. This gains us access to the quadratic mean spatial 
velocity dispersion, $3\sigma_s^2$, that in the virial theorem 
leads to the mass-to-luminosity ratio, $\Upsilon$.

\subsection{The observed velocity dispersion}

The measurement of the velocity dispersion on the observation
plane, Equation (38), needs an integration over the entire
galaxy---mathematically, over  the entire radial interval, i.e. (0, $\infty$). In
general, this condition cannot be satisfied and we  can measure only
$I(R) \sigma_p^2(R)$ in a radial interval $(0,R_m)$ which does not
contain  the entire luminosity, $L$, of the galaxy. This can lead to an
incorrect  value of $\sigma_p^2$ and therefore a wrong estimate
of the mass-to-luminosity ratio, $\Upsilon$, from the virial theorem.

 However, this lack of spatial coverage in the observations
can be compensated thanks to our galaxy model. Once we
have the density distribution law, Equation (17), and
the potential distribution, Equation (32), the spatial
velocity dispersion, $\sigma_s^2(r)$ (assuming isotropy in the
velocity distribution function), satisfies the Maxwell--Jeans 
equation,

\begin{equation}
{ d \over {\rm{d}}r } \rho(r) 3\sigma_s^2(r) = - \rho(r)
  { d \over {\rm{d}}r } \phi(r),
\end{equation}
and, with the natural boundary condition $\rho(r)\sigma_s^2(r)\rightarrow 0$
for $r\rightarrow \infty$, we have

\begin{equation}
 \rho(r) 3\sigma_s^2(r) = \int_{r}^{\infty} \rho(r') 
{d \over {\rm{d}}r'} \phi(r')  {\rm{d}}r'  =
G \int_{r}^{\infty} \rho(r') 
{M(r') \over r'^2}  {\rm{d}}r'.
\end{equation}

Equations (17) for  $\rho(r)$ and (29) for $M(r)$ allow the
direct calculation of $\sigma_s^2(r)$ from $r=0$ to 
$r=\infty$. Afterwards, we can use  Equation (37)
to get, via our model, the local mean velocity dispersion,
$\sigma_p^2(R)$, in the observation plane.

We have computed both $\sigma_s^2(r)$ from Equation (40) and
 $\sigma_p^2(R)$ from Equation (37) for each value of $n$
($n=1,2,...,10$)
in the S\'ersic profile (see Fig. 4). In all the cases we have
calculated the corresponding quadratic mean over the total 
space, i.e. $\sigma_s^2$ and $\sigma_p^2$. We have found
that ($GL/R_e$)$w^2=3\sigma_s^2$ and $\sigma_s^2=\sigma_p^2$ with
 a precision greater that the
$0.01\%$, i.e. they are satisfied to at least with four significant
number. This has been the precision that we have used
in the calculation of the density, $\rho(r)$.

 As we now have  a ``theoretical'' distribution of
$\sigma^2_p(R)$, we can compute the mean quadratic
value, $\sigma^2_p(R_m)$, inside any total radius $R_m$ and
so we can evaluate the difference between the total
and any partial integration. In Fig. 5 we show the effects that
can appear as a consequence
of having a lack of data in the radial observational interval. We
have calculated the quadratic mean partial value,

\begin{equation}
\sigma_p^2 (R_m) = { \int_{0}^{R_m} \sigma_p(R) 
 {\rm{d}}L(R)  \over \int_{0}^{R_m} {\rm{d}}L(R) }.
\end{equation}

Obviously, when $R_m \rightarrow \infty$ we find the total value 
$\sigma_p^2 = 1/3 (GL/R_e) w^2$. But when $R_m/R_e$ is too low
 we can find important differences between
$\sigma_p^2$ and $\sigma_p^2 (R_m)$, at least for some
values of $n$ in the S\'ersic profile (see Fig. 5).

\section{Conclusions}

 Once we admit that the distribution of the intensity over
a galactocentric radius of an elliptical galaxy can be
well represented by the $R^{1/n}$ S\'ersic profile a
semi-analytical expression for the corresponding spatial density of
luminous sources, $\rho_{L}(r)$, is given. It takes the
form of a sum of exponentials with the same argument,
$R^{1/n}$ ,as in the observed intensity profile,
$I(R)$. But for $\rho_{L}(r)$ in each one of these exponentials,
the argument $r^{1/n}$ is multiplied by a numerical factor,
$\lambda_j$, and is easily obtained. Likewise, the corresponding
coefficient of each exponential is easily computed. The number ($N_{ap}$)
of these exponentials, the order of approximation, depends
on the required precision. A number between 5 and 10 can
be sufficient for practical applications. Once this semi-analytical
expression for the spatial density, $\rho(r)$, is found, the distribution
of mass, $M(r)$, the potential, $\phi(r)$, and the
velocity dispersions, $\sigma_s^2(r)$ and $\sigma_p^2(R)$, can
be  computed in a straightforward manner.

 Furthermore, we show that the total quadratic mean of the measured
velocity dispersion over the observational plane, $(GL/R)\sigma_p^2(R_m)$,
cannot take the adequate value $(GL/R)\sigma_p^2$
as a consequence of an incomplete integration because of the lack of
observations. The ratio between $\sigma_p^2(R_m)$ and $\sigma_p^2$
obtained by means of computations with parameters and the function
of the ``S\'ersic'' models provides us with the corresponding
correction factor and consequently the correct value to use in the 
virial theorem to deduce the mass-to-luminosity ratio.

\section*{Acknowledgments}

  We thank Terry Mahoney and Enrique Perez for corrections to the manuscript.

\newpage

{\bf Figure Captions:}

\vspace{1cm}

{\bf Figure 1:} {\it Top:} the surface distribution of
intensity, $I(R)$; S\'ersic law for different values of 
the exponent $n$ ($n=1,2,...,10$) in  Equation (7). The
dotted line correspond to $n=1$. {\it Bottom:} the corresponding  
normalized luminosity distribution, $L(R)/R$, equation (8) and (9).

{\bf Figure 2:} {\it Top:} plot of the density of luminosity $\rho(r)$ corresponding to the $I(R)$ S\'ersic profiles for $n=1,2,...,10$. {\it Bottom:}
plot of the normalized mass $M(r)/M$ for the same values of $n$.

{\bf Figure 3:} Curves for the normalized potential $\phi(s)/\phi(0)$ for 
$n=1,2,...,10$.

{\bf Figure 4:} {\it Top:} plot of the spatial velocity dispersion $\sigma_s^2(r)$ derived from equation (40) for $n=1,2,...,10$. {\it Bottom:}  plot of the observed velocity dispersion $\sigma_p^2(R)$ derived from 
equation (37) for the same values of $n$.

{\bf Figure 5:} Plot of the aperture velocity dispersion $\sigma_p^2(R_m)$ derived from equation (41) for different values of the exponent $n$ ($n=1,2,...,10$) in the S\'ersic profile.

\bsp

\label{lastpage}

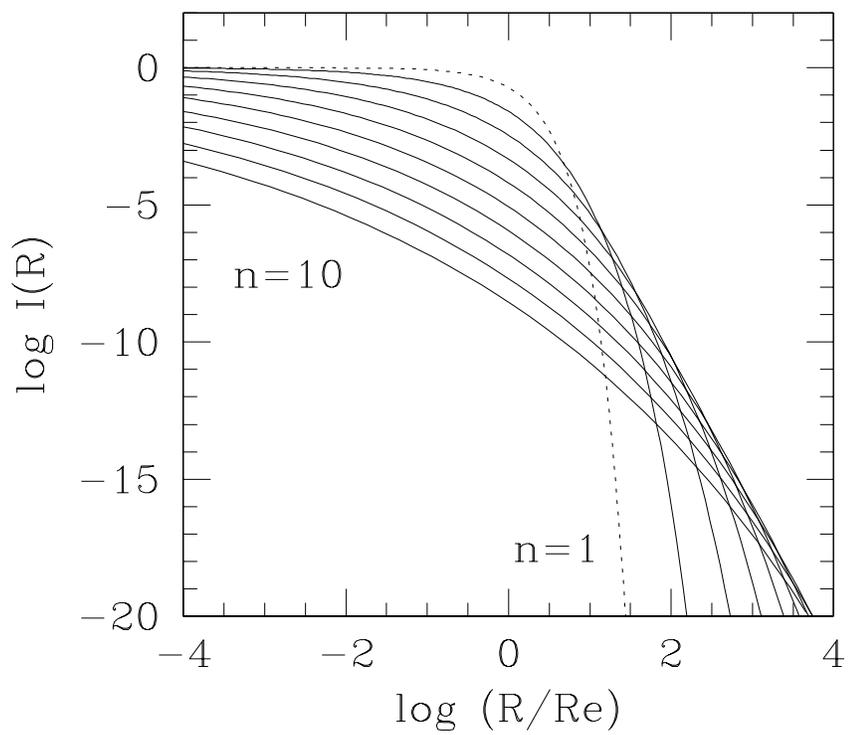
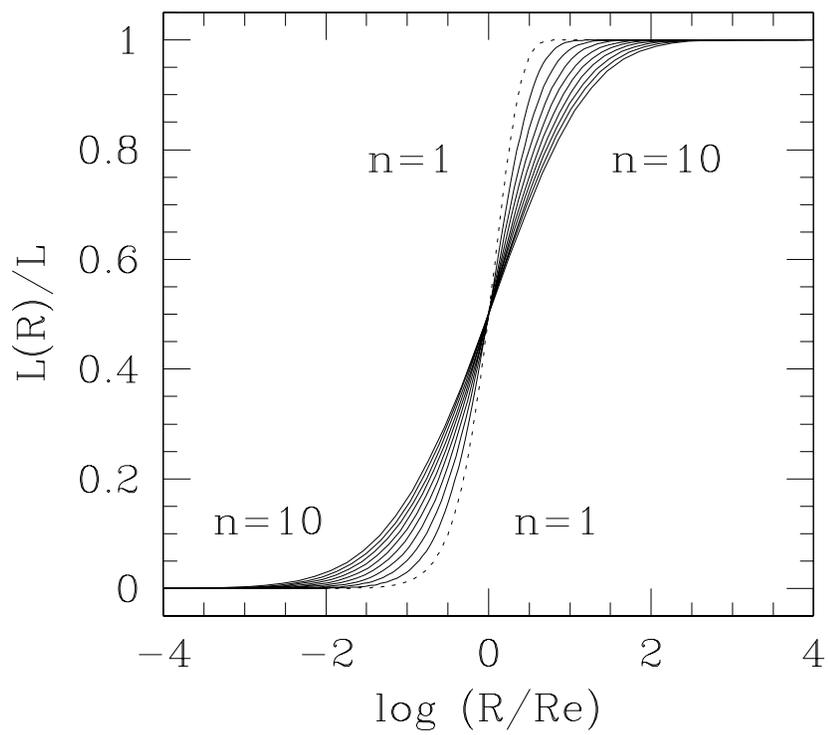
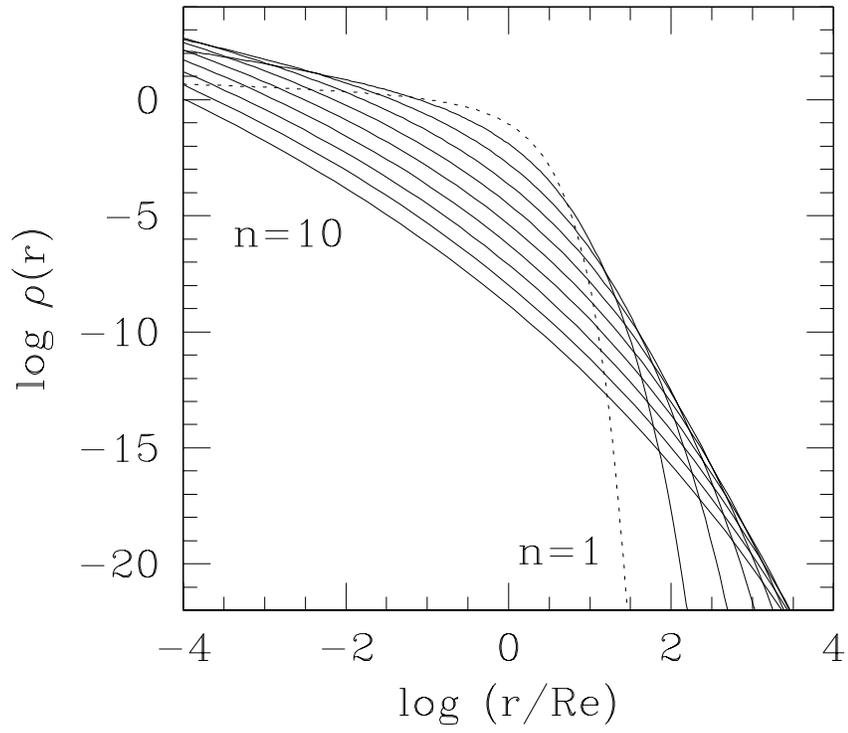
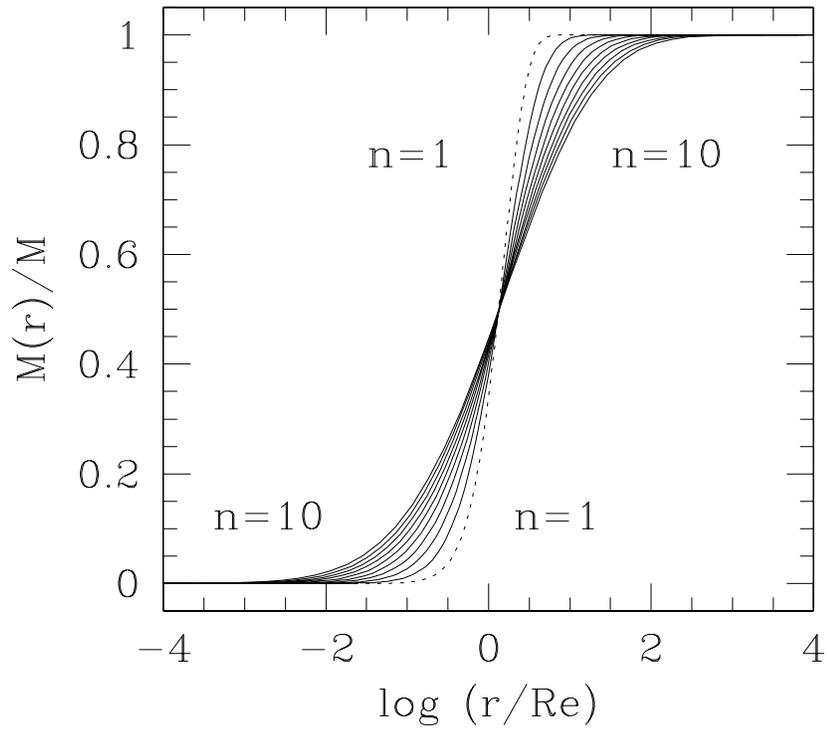
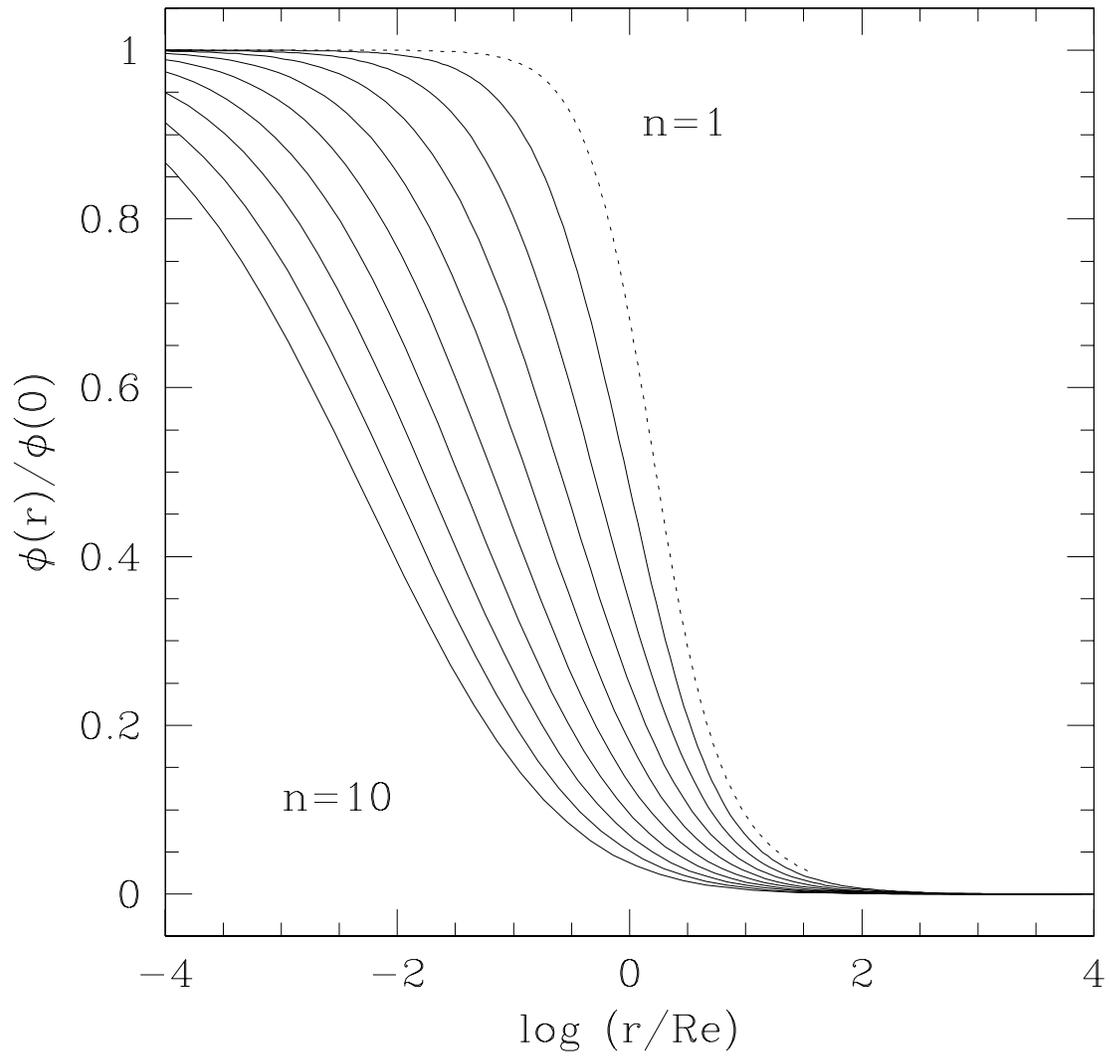
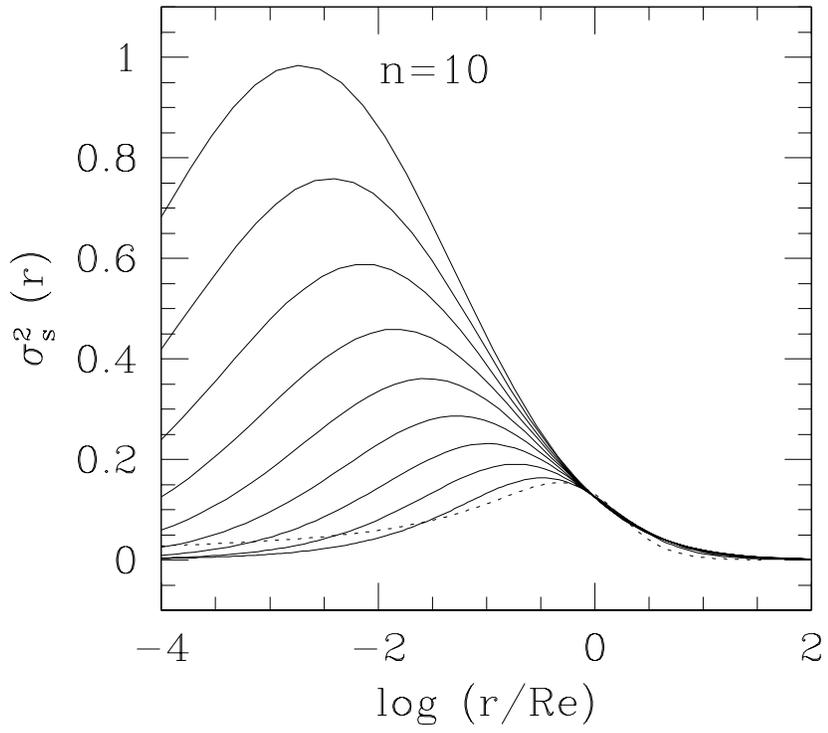
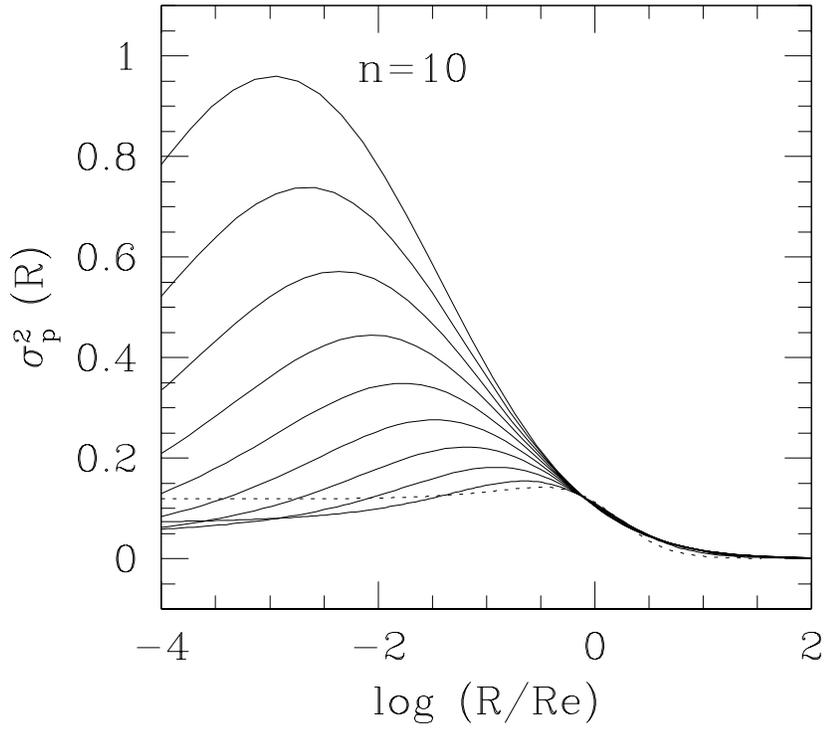
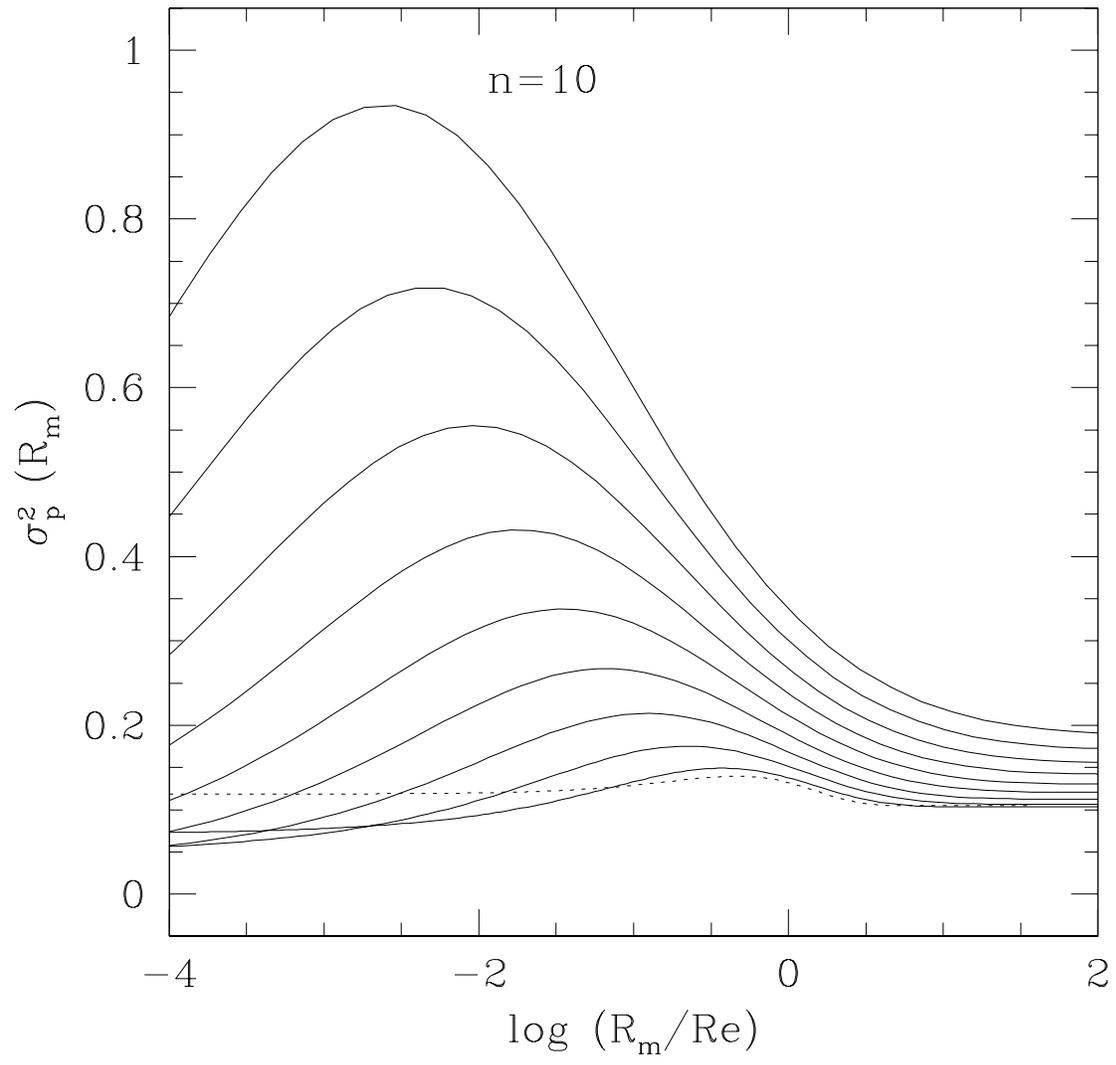
\end{document}